\documentclass[twocolumn,secnumarabic,amssymb, nobibnotes, aps, prd]{revtex4-2}
\usepackage{calc}
\usepackage{natbib}
\usepackage{graphicx}
\usepackage{dcolumn}
\usepackage{bm}
\usepackage{epstopdf, epsfig}
\usepackage{amsmath}
\usepackage{amssymb}
\usepackage{pgfplots}
\usepackage{booktabs}
\usepackage{verbatim}
\usepackage{subfig}
\usepackage{tikz}
\usepackage{dashrule}
\usepackage{comment}
\usepackage{lipsum}
\usepackage{float} 
\usepackage{xcolor}

\begin{document}
	
\title{Kolmogorov scaling in bubble-induced turbulence}

\author{Tian Ma$^{1}$}
\email{tian.ma@hzdr.de}
\author{Shiyong Tan$^{2}$, Rui Ni$^{2}$}
\author{Hendrik Hessenkemper$^{1}$}
\email{h.hessenkemper@hzdr.de}
\author{Andrew D. Bragg$^{3}$}
\email{andrew.bragg@duke.edu}

\affiliation{$^{1}$Helmholtz-Zentrum Dresden-Rossendorf, Institute of Fluid Dynamics, 01328 Dresden, Germany}

\affiliation{$^{2}$Johns Hopkins University, Department of Mechanical Engineering, MD 21218, USA}

\affiliation{$^{3}$Department of Civil and Environmental Engineering, Duke University, NC 27708, USA}

\date{\today}
	
\begin{abstract}
	
Experiments using 3D Lagrangian tracking are used to investigate Kolmogorov scaling below the bubble size in bubble-induced turbulence (BIT). Second and third order structure functions reveal approximate Kolmogorov scaling for homogeneous bubble swarms. A new scaling for the kinetic energy dissipation rate is derived and shown to be in excellent agreement with the data. Using this we predict the scale separation below the bubble size as a function of the parameters and find that a large inertial range is not possible in BIT since bubbles of the required size would quickly break down.
\end{abstract}

\maketitle

\textit{Introduction}.—Bubble driven flows are crucial in many industrial and natural process \cite{2002_Deane,2010_Balachandar,2018_Lohse,2020_Mathai,2021_Schlueter,2022_Brandt,2022_Deike,2022_Atasi,2022_Liao,2023_Ma_a}. When the bubble diameter $d_b$ is large enough, their rising motion generates turbulence in the flow, referred to as bubble-induced turbulence (BIT). The resulting dynamics becomes particularly complex as the gas void fraction increases since then, not only are the individual bubble wakes turbulent, but the diffusion of wake turbulence, wake-wake and wake-bubble interactions cause the entire flow field to become turbulent. This problem has been investigated both using experiments \cite{1991_Lance,2010_Riboux,2016_Prakash,2025_Ravisankar} and simulations \cite{2002_Bunner,2011_Roghair,2021_Ma,2021_Innocenti,2020_Pandey,2020_Ma,2020_Ma_a,Hidman_2023}.


Of fundamental interest are the scaling properties of the fluctuations in BIT and how they compare to those for single-phase turbulence. The latter were predicted by Kolmogorov \cite{1941_Kolmogorov_a} (K41 for brevity) for the inertial range where nonlinear inertial effects dominate the dynamics, in which the direct effects of energy injection and dissipation are negligible compared to the average downscale flux of kinetic energy $\varepsilon$. In this range, K41 predicts a turbulent kinetic energy spectrum $E(k)\sim O(\varepsilon^{2/3} k^{-5/3})$, where $k$ is the wavenumber.

At $k\sim O(1/d_b)$, K41 will not apply in BIT since the momentum coupling between the bubbles and flow are not accounted for in K41. However, it could apply if an inertial range exists at $k\gg 1/d_b$. Recently, direct numerical simulations (DNS) of BIT observed that when the Galilei number $Ga$ (ratio of the buoyancy to viscous forces in the flow) is large enough, $E\propto k^{-5/3}$ appears for $1/d_b <k <1/\eta$ (where $\eta$ is the Kolmogorov scale, below which viscous forces dominate) \cite{2023_Pandey}. Experimental studies, however, report conflicting results. Riboux et al. \cite{2010_Riboux} observed $E\propto k^{-3}$ for $k< 1/d_b$, and K41 scaling $E\propto k^{-5/3}$ at $k> 1/d_b$. The scaling $E\propto k^{-3}$ was previously observed by Lance \& Bataille \cite{1991_Lance} who argued that it arises due to a balance between local (in scale) energy production and dissipation. However unlike \cite{2010_Riboux}, Lance \& Bataille \cite{1991_Lance} did not observe an additional K41 range where $E\propto k^{-5/3}$.

In this letter, we explore K41 scaling in BIT using experiments equipped with state-of-the-art 3D Lagrangian particle tracking (3D-LPT) \cite{2020_Tan}. This provides unparalleled access to flow information compared to previous experimental approaches (e.g. single-point measurement \cite{1991_Lance}, 2D particle image velocimetry \cite{2010_Riboux}). For example, our approach enables measurements of the fluctuations both in the vicinity of the bubbles, and in their wake. In addition, we also derive and test a new scaling for $\varepsilon$, which together with K41, enables quantitative predictions for the properties of fluctuations below the bubble scale based on known parameters of the system.

\begin{figure}
	\centering
	\includegraphics[width=7.9cm]{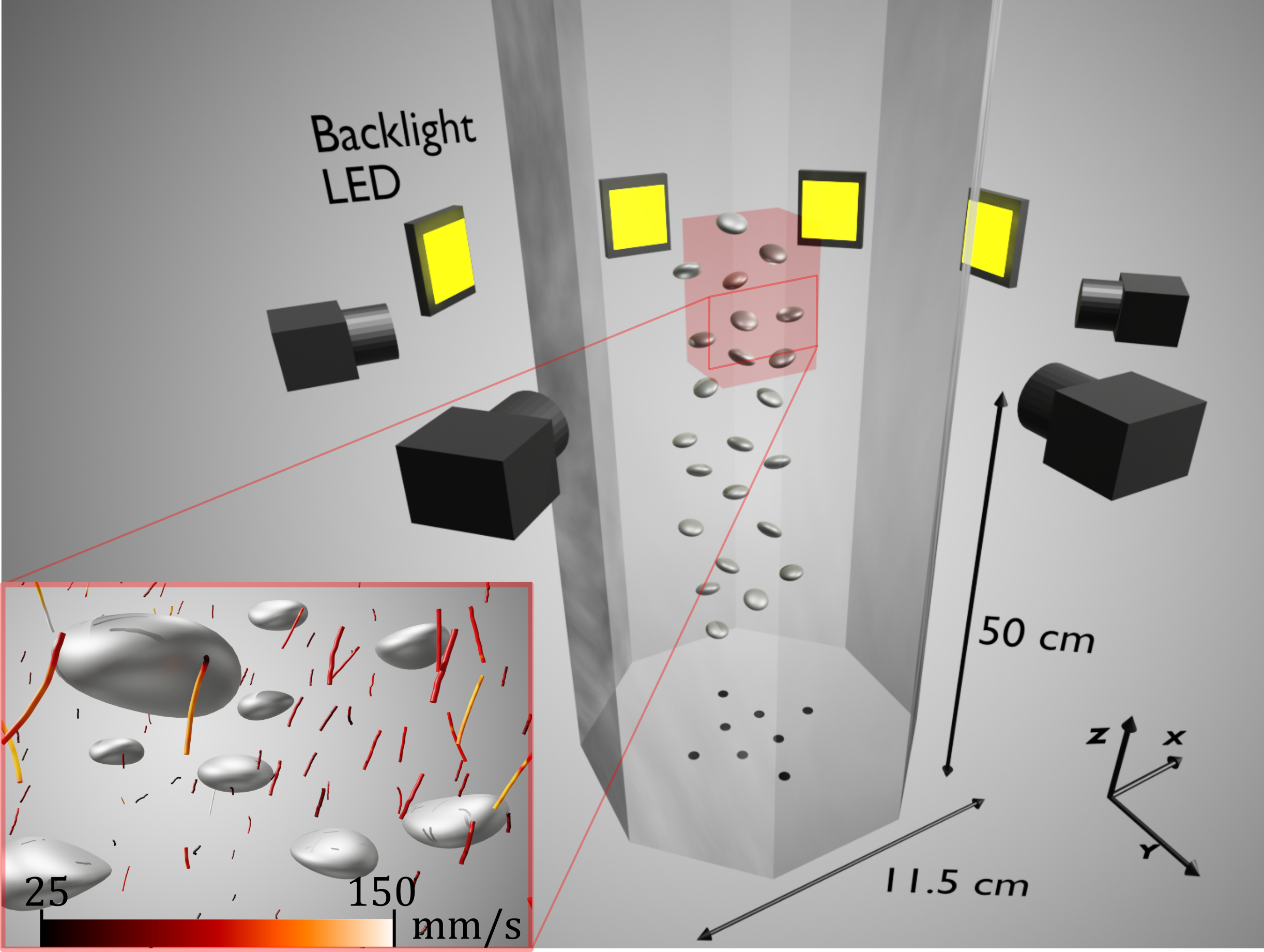}
	\caption{Sketch of the octagonal bubble column used in the experiments (note that in the actual experiment, the number of bubbles in the column is $O(10^3)$ and approximately 50 to 100 bubbles are tracked in the field of view). Inset: Sample of reconstructed 3D tracer tracks around multiple bubbles at one time instant in the \textit{Sm} case; instantaneous vertical velocity of the tracers is color-coded.}
	\label{fig: bubble column}
\end{figure}
\begin{table}
	\begin{center}
		\def~{\hphantom{0}}
		\begin{tabular}{ccccc}
			Parameter  &\textsl{Sm}&\textsl{Mid}&\textsl{La}&\textsl{LaMore}\\
			\midrule
			$d_b (mm)$      &3.5&4.1&4.4&4.6\\
			$\alpha$   &0.52\%&0.78\%&1.2\%&1.6\%\\
			$Ga$       &645&818&910&972\\
			$Eo$	   &1.7&2.3&2.6&2.9\\
			$Re_b$     &847&923&925&967\\
			{$Re_\lambda$}     &61&69&97.5&101\\
			{$Re^{out}_\lambda$}     &67&74&107&109\\
			{$d_b/\eta$}    &23&32&39&43\\
			$u^{in}_{rms}/u^{out}_{rms}$  &1.6&1.5&1.2&1.2\\
		\end{tabular}
		\caption{Parameters for the 4 monodispersed cases labeled $Sm, Mid, La$ and $LaMore$, respectively. Here, $\alpha$ the gas void fraction, $d_b$ the bubble diameter, $Ga\equiv\sqrt{gd_{b}^{3}}/\nu$ the Galileo number, $Eo\equiv(\rho_l-\rho_g)gd_b^2/\sigma$, the E\"{o}tv\"{o}s number ($\sigma$ is the surface tension), and $Re_\lambda=\sqrt{15/(\nu \varepsilon)}u^2_{rms}$ the Taylor-Reynolds number based on all the tracer tracks. $Re^{in}_\lambda$ and $Re^{out}_\lambda$ are the Taylor-Reynolds numbers calculated from tracer tracks inside and outside the bubble wakes, respectively. The bubble Reynolds number $Re_b=d_bU_b/\nu$, with the bubble rise velocity $U_b$ evaluated by subtracting the local vertical liquid velocity using the tracers within a search radius $2d_b$ from the center of a bubble. $d_b/\eta$ is the ratio of the bubble diameter to the Kolmogorov scale, and $u^{in}_{rms}/u^{out}_{rms}$ is the ratio of the r.m.s. velocity fluctuations in the wakes to those outside the wake.}
		\label{tab:database}
	\end{center}
\end{table}

\textit{Experiment and results}.—We study BIT in an octagonal bubble column (Fig. \ref{fig: bubble column}) with diameter 11.5 cm that is filled with tap water to a height of 90 cm. Bubbles are injected through 8 needle spargers inserted in the bottom. We investigate four different cases of monodisperse, homogeneous bubbly flows obtained by varying the needle size and gas flow rate. Some of the important bubble properties are listed in Table \ref{tab:database}. The bubbles are relatively large ($3\sim5$ mm, and thus in the wobbling regime \cite{1997_Lunde}) so that a considerable amount of turbulence is generated, and to help increase the scale separation between $d_b$ and $\eta$ so that a K41 regime may emerge. The gas void fraction $\alpha$ is chosen to keep all cases in the homogeneous bubbly flow regime, while also ensuring sufficient interactions between bubble wakes. For our large bubbles, $Ga$ is close to that in the DNS study \cite{2023_Pandey}, and our values for the Taylor-Reynolds number $Re_\lambda$ are also similar. 

The data is recorded in a region $4.5\,\mathrm{cm}\times4.5\,\mathrm{cm}\times9\,\mathrm{cm}$ with four calibrated high-speed cameras at an acquisition rate of 2500 frames per second over a period of 18 seconds. The bottom of the region is at height $z = 50$ cm which ensures that bubbles entering the measurement region have no memory of the way they were injected. Combining our in-house 3D Lagrangian bubble tracking tool \cite{2024_Hessenkemper} with the dense particle tracking algorithm, OpenLPT \cite{2020_Tan}, we simultaneously track the bubbles and tracers in 3D (see the illustrative movies in SM, Sec. I for 3D bubble and tracer tracks, respectively). For the latter, we use 100 $\mu$m polyamide seeding particles. The inset of Fig. \ref{fig: bubble column} shows representative tracer trajectories around bubbles in the \textit{Sm} case. To our knowledge, this is the first 3D simultaneous Lagrangian measurement for both phases, considering dense deformable bubbles. More details on the experimental methods can be found in SM, which
includes Refs. \cite{2020_Tan,2016_Schanz,2023_Schroder,2022_Kim,2021_Salibindla,2023_Tan,2022_Hessenkemper,2024_Hessenkemper,2016_Cano,2024_Liu,2025_Xu,Wang_2025}.


Standard approaches for computing $E(k)$ require a uniformly spaced flow signal, and this is problematic in BIT because the bubbles interfere with the measurements. Indeed, for this reason, in Ref.~\cite{1991_Lance} a Gaussian smoothing function was used to smooth the flow signal across points where it was disrupted by a bubble. Furthermore, spectrums only provide limited information on the flow. Therefore, in order to more comprehensively analyze the flow, we consider the \textit{n}th-order structure functions $\boldsymbol{D}_n(\boldsymbol{r},t)\equiv\langle \Delta \boldsymbol{u}^n(\boldsymbol{x},\boldsymbol{r},t) \rangle$, where $\Delta \boldsymbol{u}(\boldsymbol{x},\boldsymbol{r},t)\equiv \boldsymbol{u}(\boldsymbol{x}+\boldsymbol{r},t)-\boldsymbol{u}(\boldsymbol{x},t)$ is the fluid velocity increment between two points in the flow separated by $\bm{r}$, and $\left\langle \cdot \right\rangle$ denotes an ensemble average (the flow in the measurement region is approximately homogeneous, hence $\boldsymbol{D}_n$ is independent of $\boldsymbol{x}$). Projections of $\boldsymbol{D}_2, \boldsymbol{D}_3$ along $\bm{r}$ yield the longitudinal components $D_{LL}, D_{LLL}$, respectively, while projections orthogonal to $\bm{r}$ yield the transverse components $D_{TT},D_{TTT}$.

\begin{figure}
	\centering
	\includegraphics[height=5cm]{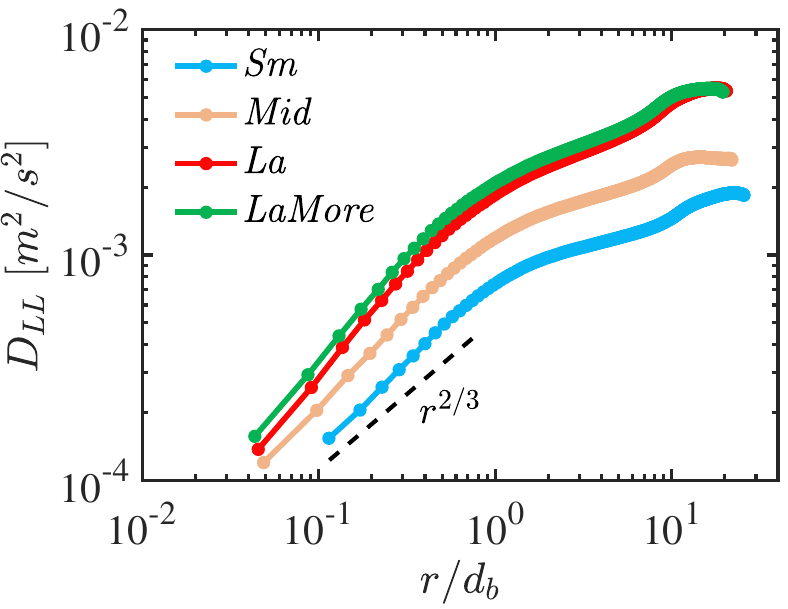}
	\caption{Longitudinal second-order structure functions. The black dashed lines indicate slope $r^{2/3}$, evident at the small scales for $r/d_b<1$.} \label{fig: SF}
\end{figure}

In the inertial range, K41 predicts $D_{LL}=(3/4)D_{TT}\sim O([\varepsilon r]^{2/3})$, and in Fig.~\ref{fig: SF} we test these scaling predictions against the experimental data. For \emph{Sm} and \emph{Mid}, $D_{LL}$ and $D_{TT}$ approximately follow $r^{2/3}$ over almost one decade of $r/d_b$ for $r<d_b$, with a slight bump at $r\sim d_b$. In the context of 3D measurements taken in the vicinity of the bubble swarm, these results provide the first experimental confirmation of the DNS results in \cite{2023_Pandey} that K41 scaling can apply in BIT at scales $<d_b$.  For the larger bubble cases $D_{LL}$ and $D_{TT}$ are steeper than $r^{2/3}$ at $r<d_b$. This may be because the larger bubbles are approaching the heterogeneous BIT regime for which a mean shear is generated that is not accounted for by K41. 

\begin{figure}
	\centering
    \makebox[1.3em][l]{\raisebox{-\height}{(\textit{a})}}%
	\raisebox{-\height}{\resizebox{0.75\linewidth}{!}{\includegraphics{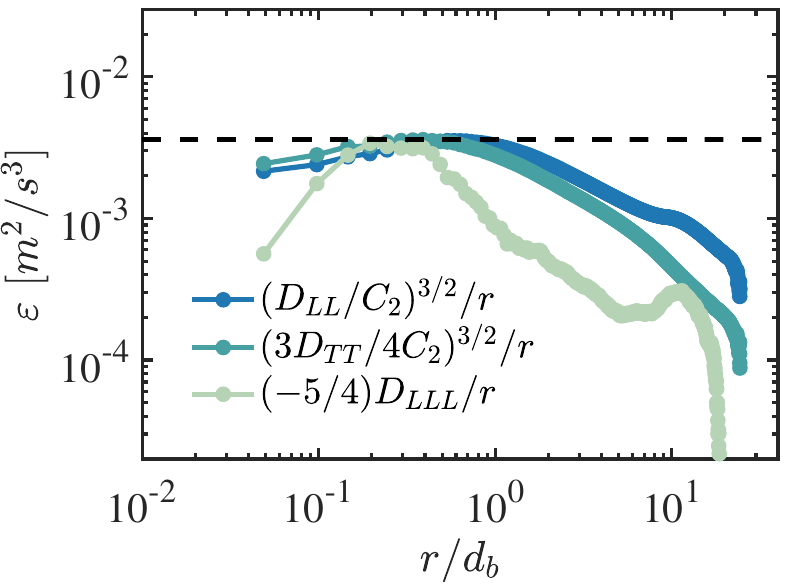}}}
	\centering
    \makebox[1.3em][l]{\raisebox{-\height}{(\textit{b})}}%
	\raisebox{-\height}{\resizebox{0.75\linewidth}{!}{\includegraphics{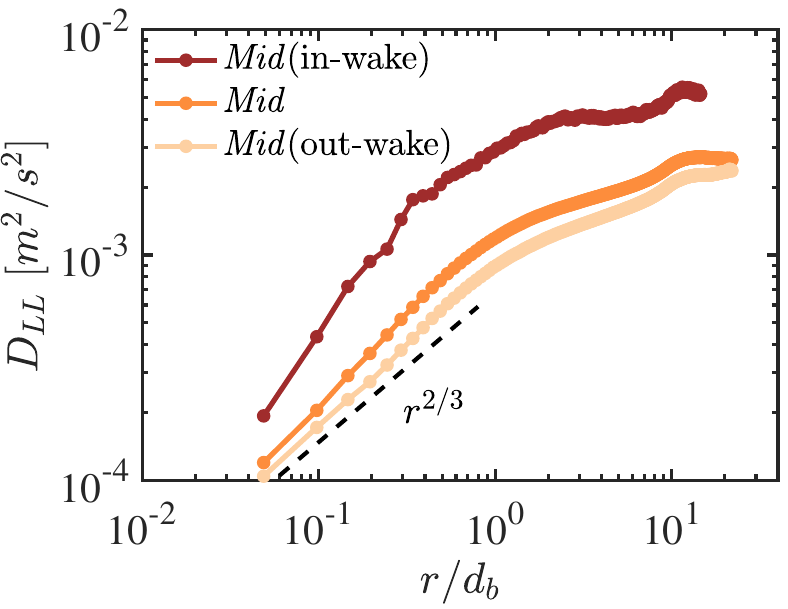}}}
	\caption{(\textit{a}) Turbulent energy dissipation rate for the \textit{Mid} case, estimated by structure functions compensated by their respective scaling laws. The black dashed line is used to estimate $\varepsilon$. (\textit{b}) Longitudinal second-order structure functions for the \textit{Mid} case, calculated using tracer tracks inside the wake, all tracer tracks and tracks outside the wake, respectively. The black dashed line indicates slope $r^{2/3}$.}
	\label{fig: dissipation}
\end{figure}

Although the data for $D_{LL}$ and $D_{TT}$ is consistent with K41 for $r<d_b$ for the \emph{Sm} and \emph{Mid} cases, this is not sufficient to confirm the existence of an inertial cascade and true K41 scaling. To further test for this we can consider the ``four-fifths law'' of Kolmogorov for the third-order structure function $D_{LLL}=-(4/5)\varepsilon r$ (which although consistent with K41, can be derived from the Navier-Stokes equation \cite{1941_Kolmogorov_b}). If K41 applies at $r<d_b$ we should expect not only that the data here agrees with $D_{LLL}=-(4/5)\varepsilon r$, but also that $(D_{LL}/C_2)^{3/2}/r=(3 D_{TT}/4C_2)^{3/2}/r=-(5/4)D_{LLL}/r$, with their common value being equal to $\varepsilon$. Figure~\ref{fig: dissipation}(\textit{a}) shows the results for these quantities for a representative case \textit{Mid} (other cases are similar, see SM, Sec.II). The results confirm quite well that K41 scaling holds for $D_{LL}$ and $D_{TT}$, with the two curves almost collapsing and plateauing to the same constant value for $0.2\lesssim r/d_b\lesssim1$ \footnote{That the curves begin to decay below $r/d_b\lesssim 0.2$ is most likely due to the dissipation scale being approached. In isotropic turbulence, the scaling gradually steepens from $r^{2/3}$ to $r^2$ as $r/\eta$ reduces towards the regime $r/\eta\leq O(1)$. Similarly, the decay of the curves for $r/d_b\lesssim 0.2$ is likely due to the scaling $r^2$ starting to be approached, although it is not actually attained in our data.}, suggesting $\varepsilon\approx0.0036$ $m^2/s^3$ for this case. Note that in these plots we use the value $C_2=2.1$ from single-phase turbulence; remarkably the same value seems to describe BIT well. The two curves are however not exactly constant for $0.2\lesssim r/d_b\lesssim1$, and $(D_{LL}/C_2)^{3/2}/r$ is slightly smaller than $(3 D_{TT}/4C_2)^{3/2}/r$ for $r<d_b$, indicating weak anisotropy of the fluctuations (under K41's assumption of local isotropy, these curves would be identical). These deviations are likely due to the fact across the cases we have $d_b/\eta\in[23,43]$, while true K41 scaling would require much larger scale separations. The curve for $(-5/4)D_{LLL}/r$ also becomes approximately constant at $r<d_b$ but over a smaller range, $0.2\lesssim r/d_b\lesssim 0.5$, indicating that this is not a well-developed inertial range. This is again not surprising since in single-phase isotropic turbulence $(-5/4)D_{LLL}/r$ does not approach a constant value over an extended range of scales until $Re_\lambda$ is large enough for a wide separation of scales in the flow \cite{2009_Ishihara}. 
Estimating $\varepsilon$ from these curves, we can then calculate several important parameters that are traditionally used to characterize turbulence: (i) the Kolmogorov scale of the BIT flow $\eta=(\nu^3/\varepsilon)^{1/4}$. (ii) The Taylor-scale Reynolds number: $Re_\lambda=\sqrt{15/(\nu \varepsilon)}u^2_{rms}$, where $u_{rms}$ is the root mean square of the fluctuating liquid velocity. Their values for all the cases are shown in Table \ref{tab:database}. As expected, $Re_\lambda$ increases from \textit{Sm} via \textit{Mid} to \textit{La/LaMore}, reflecting an increased intensity of BIT. In isotropic turbulence, such values (e.g. $Re_\lambda\approx101$ for \textit{LaMore}) would be classified as moderately turbulent, and can be generated in wind/water tunnels \cite{2010_Monchaux,2018_Mathai}. It is surprising that purely bubble-driven flows are able to generate turbulence with this intensity given that the void fraction in the present study is quite small $\alpha=1.6\%$. 

We can utilize our detailed data of the flow to investigate the properties of the flow inside and outside of the wake regions (see SM, Sec.~III for how to extract the corresponding tracer tracks in these two regions). Figure~\ref{fig: dissipation}(\textit{b}) again takes \textit{Mid} as a representative case and shows that K41 scaling applies very well to $D_{LL}$ and $D_{TT}$ outside the wake regions, and in fact slightly better than when all regions of the flow are included (see $D_{TT}$ in SM). The scaling observed inside the wake regions at $r<d_b$ is significantly steeper than the K41 prediction. This is perhaps due to the turbulence generation mechanism (shear stresses generated at the bubble surfaces) being too strong in these regions for the nonlinear inertial effects to become dominant enough for K41 to apply. It could also be due to the strong flow anisotropy in these regions, which may violate the assumption of local isotropy in K41. 

At all scales $D_{LL}$  and $D_{TT}$ are much larger inside the wakes than outside. This is simply due to the fact that the turbulence intensity is strongest in the wakes since these are the regions closest to where the turbulent fluctuations are produced by stresses on the bubble surfaces. Associated with this is that the r.m.s. velocity fluctuations in the wakes $u^{in}_{rms}$ are larger than those outside $u^{out}_{rms}$ (see Table \ref{tab:database}). The fluctuations outside of the wake are, however, intense enough to generate sufficient inertia for K41 scaling, and in fact $Re_\lambda$ is slightly larger outside of the wakes than when all regions of the flow are included, since the Taylor length scale is larger in the former than the latter. This is likely due to mixing of the wake region which causes the flow length scales to grow with time. Furthermore the ratio $u^{in}_{rms}/u^{out}_{rms}$ decreases with $Re_\lambda$ (see Table \ref{tab:database}), indicating that the flow becomes more homogeneous and well mixed from the case \textit{Sm} via \textit{Mid} to \textit{La}/\textit{LaMore}. At sufficiently low $\alpha$ one would not expect to see K41 scaling outside of the wakes because in this case most of the flow will be almost quiescent.

\begin{figure}
	\centering
	\includegraphics[width=6.5cm]{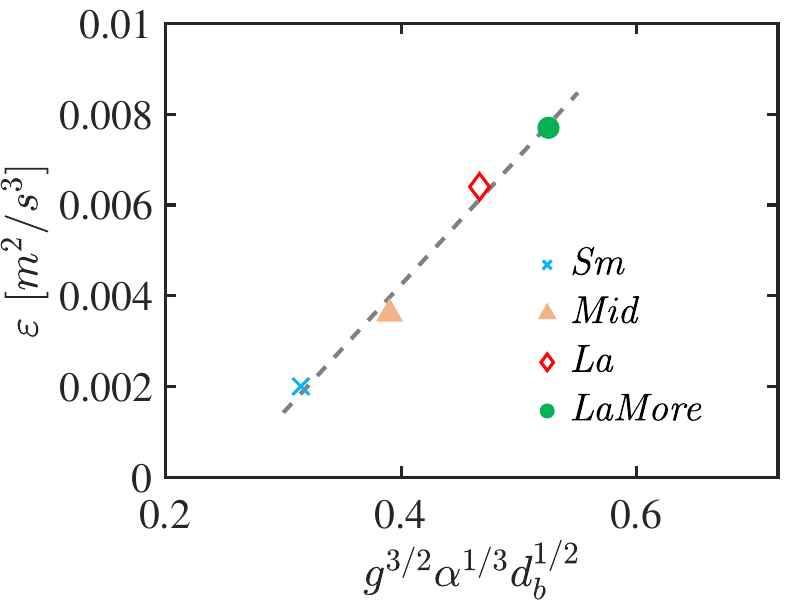}
	\caption{The dependence of the turbulent kinetic energy dissipation rate $\varepsilon$ on $g^{3/2}\alpha^{1/3}d_b^{1/2}$. The dashed line indicates the linear relationship predicted by \eqref{eq: dissipation}.}
	\label{fig: dissipation model}
\end{figure}

In order to make quantitative predictions of BIT at $r<d_b$ using K41 requires estimating $\varepsilon$ in terms of known parameters of the flow. Dimensional analysis and the assumption of an energy cascade below the bubble scale suggests $\varepsilon\sim O(U^3/D)$, where $D$ and $U$ are relevant length and velocity scales. The choice $D= d_b$ is not suitable since although $d_b$ plays a crucial role in determining the wake size, BIT also depends on the spacing between the bubbles which determines wake-wake interactions. The latter is influenced by $\alpha$, and this suggests that a more appropriate choice is the average inter-bubble distance $D=d_b/\alpha^{1/3}$, which is also on the order of the bubble wake length according to the data in \cite{Zamansky_2024}. This is analogous to single-phase grid-generated turbulence where the grid spacing sets the integral scale $L$ which in turns determines $\varepsilon$ according to $\varepsilon\sim O(u_{rms}^3/L)$ \cite{1974_Gad}. For $U$ an appropriate choice is the mean bubble rise velocity $U=U_b\sim O(\sqrt{gd_b})$, and with these choices we obtain
\begin{equation}
	\varepsilon \sim C g^{3/2}\alpha^{1/3}d_b^{1/2},
	\;\label{eq: dissipation}
\end{equation}
where $C$ is independent of $g,\alpha,d_b$, but may depend on other properties of the system. In Fig. \ref{fig: dissipation model} the values of $\varepsilon$ obtained from the experiments are plotted as a function of $g^{3/2}\alpha^{1/3}d_b^{1/2}$, and the dashed line indicates the linear relationship predicted by (\ref{eq: dissipation}). The agreement is excellent across the cases (although a more thorough test would require independently varying $\alpha$ and $d_b$, which is not possible with our data), and since only two parameters ($\alpha$ and $d_b$) are required by this scaling to estimate $\varepsilon$, it provides, together with K41 scaling, simple algebraic expressions that accurately predict key flow properties in BIT. The analysis in \cite{Zamansky_2024} assumes that $\varepsilon$ is determined by viscosity and the wake-produced shear, rather than by an energy cascade as we have assumed, and this leads to a very different scaling $\varepsilon \sim \nu\alpha^{2/3}g/d_b$. We cannot demonstrate which model for $\varepsilon$ is most reasonable since in the experiments $g,\nu$ are fixed and the range of $\alpha, d_b$ is limited. We note, however, that while our model is consistent with the K41 scaling we observe for the structure functions, the viscous scaling of \cite{Zamansky_2024} is not.

Using the model for $\varepsilon$ we obtain
\begin{align}
d_b/\eta\sim C^{1/4} d_b^{9/8}g^{3/8}\alpha^{1/12}\nu^{-3/4}=C^{1/4} Ga^{3/4}\alpha^{1/12},\label{scale_range}
\end{align}
where the data indicates $C^{1/4}\approx 0.42$, and using \eqref{scale_range} we can estimate the ranges of $d_b$ and $\alpha$ that would be required in order to observe a well-developed inertial range in BIT. In isotropic turbulence, inertial scaling begins to clearly develop when $L/\eta\gtrsim 200$ \cite{DONZIS_2005}, where $L$ is the energy injection scale. Since the energy injection scale in BIT is $O(d_b)$ we can then estimate that clear inertial scaling will occur in BIT when $d_b/\eta\gtrsim 200$. According to \eqref{scale_range}, $d_b/\eta$ is weakly dependent on $\alpha$ due to the $1/12$ exponent, and its main dependence is on $d_b^{9/8}$ (considering $g,\nu$ fixed). Table \ref{tab:database} shows the current values of $d_b/\eta$ for each case, and using \eqref{scale_range} we can estimate how much larger $d_b$ would need to be in each case to achieve $d_b/\eta= 200$ (with $\alpha$ fixed since the dependence on $\alpha$ is weak). For cases with values of $\alpha$ corresponding to those in \textit{Sm}, \textit{Mid}, \textit{La} and \textit{LaMore} we would need bubbles of size $d_b\approx 29$ mm, $25$ mm, $21$ mm and $20$ mm, respectively. In the presence of turbulence, bubbles of such size would quickly break up \cite{2024_Ni}. Hence, there is an intrinsic limitation in BIT regarding the size of the inertial range possible, and how accurately K41 can apply. 

The argument above suggests that larger $d_b/\eta$ are required for K41 to accurately apply in BIT. However, $d_b/\eta$ is almost twice as large for \textit{LaMore} than \textit{Sm}, yet Fig. \ref{fig: SF} shows that K41 performs better for \textit{Sm} than \textit{LaMore}. As indicated earlier, this is probably because the larger bubbles are approaching the heterogeneous BIT regime, to which K41 is less applicable.

\textit{Discussion}.—The observation of approximate K41 scaling below the bubble scale helps to resolve the controversial question regarding the applicability of K41 scaling in BIT. Our results are in agreement with DNS results in \cite{2023_Pandey}, and what both studies have in common is that $Ga$ is relatively large, $Ga\gtrsim O(1000)$. DNS studies \cite{2020_Pandey,2021_Innocenti,Hidman_2023} did not observe K41 scaling, but reported $E\propto k^{-3}$ below the bubble scale. This could be due to the lower values of $Ga$ in those studies; $Ga_{max}=358$ in \cite{2020_Pandey}, $Ga=185$ in \cite{2021_Innocenti} and $Ga=390$ in \cite{Hidman_2023}. Indeed, Pandey et al. \cite{2023_Pandey} noted that $E\propto k^{-5/3}$ only emerges in their DNS for $Ga=O(1000)$, and that at lower $Ga$ the scaling $E\propto k^{-3}$ dominates. That K41 scaling only emerges for sufficiently large $Ga$ is consistent with the scaling result in \eqref{scale_range}. Several things remain unclear, however. Why did \cite{1991_Lance} not observe K41 scaling below the bubble scale despite the fact that $Ga=O(1000)$? This could be due to limitations associated with the use of Taylor's hypothesis to convert their time-series data into spatial data for computing $E(k)$. The frequency-space results in \cite{2016_Prakash} did not show K41 scaling below the bubble scale despite having large $Ga$. However, this could simply highlight issues associated with applying K41 to frequency-space variables, just as it faces challenges when applied to Lagrangian variables in isotropic turbulence \cite{Biferale08}. Why did \cite{2010_Riboux} observe K41 despite having values in the range $Ga\in[200,391]$ for which DNS did not observe K41? This could be due to differences in the experimental and DNS flows, e.g. the use of periodic boundaries in the latter. Moreover, an important difference between the DNS/our experiments and the experiments of \cite{2010_Riboux} is that while results from the DNS/our experiments are based on measurements in the midst of the bubble swarm, measurements in \cite{2010_Riboux} were only taken after all the bubbles had risen past the measurement location.

Of the studies reporting a K41 scaling range, there is some disagreement regarding the coexistence of a separate $E\propto k^{-3}$ range. The DNS study \cite{2023_Pandey} claimed to observe $E\propto k^{-3}$ at scales below $\eta$. However, we note that this behavior appears to hold only over a limited range of wavenumbers, which limits the evidential support for the scaling regime. The study of \cite{2010_Riboux} claimed, on the other hand, to observe $E\propto k^{-3}$ above the bubble scale. A scaling $E\propto k^{-3}$ should correspond to $D_{LL}\propto D_{TT}\propto r^2$, which we do not observe in our experiments above the bubble scale (nor is any clear scaling observed in this range). It is vital that future work seeks to resolve all of these discrepancies among existing studies.


\textit{Conclusion}.—Our results provide the first experimental evidence, based on 3D measurements of the flow, that at scales below the bubble diameter $d_b$, approximate K41 scaling can emerge. The K41 scaling observed in our experiments is not perfect (nor was it in \cite{2023_Pandey}), and this is due to the limited range of scales below $d_b$. Using our scaling for $\varepsilon$ we concluded that a well-developed inertial range (required for precise K41 scaling) cannot emerge in BIT because bubbles of the required sizes would quickly break down. Our findings enrich the class of out-of-equilibrium systems exhibiting approximate Kolmogorov scaling, along with recent discoveries for porous media \cite{2024_Falkinhoff} and active matter \cite{2020_Bourgoin} flows.

\textit{Acknowledgments}.—T.M. thanks G. Huang for support during the revision. The authors acknowledge financial support from the German Research Foundation (DFG) under grant number 536223645.

\bibliography{BIT_exp_PRL}

\end{document}